\documentclass[
    ,final            
  ]
  {aipproc}

\layoutstyle{6x9}

\begin{document}

\title{On the S-Matrix of the Faddeev-Reshetikhin Model}

\classification{02.30.Ik,11.55.Ds,11.10-z}
\keywords      { Sigma Models, Integrable Field Theories, Bethe Ansatz, Exact S-Matrix}

\author{Victor O. Rivelles}{
  address={Instituto de Física, Universidade de São Paulo, 05315-970, São Paulo, SP, Brazil}
}

\begin{abstract}
 The Faddeev-Reshetikhin model arises as a truncation of strings in $AdS_5
\times S^5$. Its two particle S-matrix should be obtained by diagonalizing its
Hamiltonian. However this does not happen in a straightforward way. There is a
Lorentz violating term in the Hamiltonian which prevents its plain
diagonalization. We then find out the most general quartic Hamiltonian that can
be diagonalized. It includes the bosonic Thirring model as well as another model
which shares the same two particle S-matrix as the Faddeev-Reshetikhin model. We
also find an one parameter family of  interactions which lead to non-Hermitian 
Hamiltonians which have unitary S-matrices and are PT symmetric.
\end{abstract}

\maketitle


\section{Introduction}

Ten years ago a daring conjecture \cite{Maldacena:1997re,Gubser:1998bc,Witten:1998qj} was made between two apparently disconnect theories. It was proposed that string theory in $AdS_5 \times S^5$ is dual to the ${\cal N}=4$ supersymmetric gauge theory in four dimensions. Ten years later \cite{Aharony:1999ti,Minahan:2006sk}, a huge amount of evidence now supports that claim and it seems we are heading to a proof of the conjecture. To show that the correspondence is true it was initially thought that we  should  construct all the correlation functions of the gauge theory and all the states of the string theory. This is a huge task but these were the first steps taken to show the conjecture. Even though the relation between the two theories is a duality between the weak and strong coupling regimes many successful results were obtained \cite{Aharony:1999ti}. Reliable computations can be done in the weak coupling regime of both theories but the interpolating region was always difficult to be accessed. In the last five years, however, it was found that both theories have integrable sectors. The classical action of string theory in $AdS_5 \times S^5$ is integrable in the sense that an infinite set of non-local currents can be found \cite{Bena:2003wd}. On the other side, the problem of finding the anomalous dimensions of gauge invariant operators in the planar limit of the gauge theory can be mapped to the problem of diagonalizing a spin chain \cite{Minahan:2002ve}. As a result, the energy spectrum of the string is conjectured to coincide with the spectrum of anomalous dimensions. Thanks to the integrability properties of both sides this was verified in several limits where they can be compared with amazing success \cite{Minahan:2006sk}. Today we are here to celebrate the successes reached in these ten years. 

String theory is integrable at the classical level but quantum integrability is still an open issue. One approach to treat the quantum case is to consider truncated strings in $AdS_5 \times S^5$. Of course, the quantization of truncated strings can at most be regarded as a laboratory to test its integrability properties because the frozen degrees of freedom are not being properly taken  into account. Surprisingly many sectors still show integrability at the quantum level. For instance, if the string is truncated to the $R \times S^3$ sector of $AdS_5 \times S^5$, with $R$ being the time in $AdS_5$ and $S^3$ a subspace of $S^5$, then the resulting model is an $SU(2)$  principal chiral sigma model which was analysed in detail by Faddeev and Reshetikhin \cite{Faddeev:1985qu}. They proposed that the Hamiltonian for the original sigma model as well as the nonultralocal Poisson brackets should be replaced by another Hamiltonian and a new Poisson
bracket structure which lead to the same dynamical equations. After the quantization of this new system the original sigma model can be recovered by taken a certain limit. More recently, Klose and Zarembo \cite{Klose:2006dd} used standard field theoretical techniques to obtain its two particle S-matrix. If the false vacuum is chosen the calculations are strongly simplified since only bubble diagrams contribute to the S-matrix. 

An alternative way to obtain the S-matrix for the Faddeev-Reshetikhin model is by a straight diagonalization of its Hamiltonian \cite{Das:2007tb}. Amazingly, the quartic Hamiltonian for
the Faddeev-Reshetikhin model cannot be diagonalized. The trouble is in a Lorentz violating
term in the interaction Hamiltonian. This is very surprising since the calculation of the S-matrix by Klose and Zarembo \cite{Klose:2006dd} is straightforward. Even so, our results are consistent with the field theory results if we consider the Hamiltonian matrix elements of positive energy states only. This is completely consistent with the field theoretic calculation since it involves only the calculation of matrix elements. This leads to the conclusion,  which had not been observed earlier in other integrable models, that while the diagonalization of a system leads to its S-matrix, having the S-matrix does not automatically imply diagonalizability of the system.

The lack of diagonalizability of the quartic Hamiltonian then lead us to determine the most general quartic Hamiltonian which can be diagonalized assuming integrability \cite{Das:2007tb}. It includes the bosonic Thirring model, which is known to be integrable, as well as the bosonic chiral Gross-Neveu model, which to the best of our knowledge had
not been studied earlier. Both models respect Lorentz invariance and, in
fact, we
find that the two systems share the same S-matrix. We show that this puzzling coincidence 
is due to a Fierz transformation so that both models are in fact equivalent. We also determined the most general quartic Hamiltonian, violating Lorentz invariance, which can be diagonalized and leads to the same S-matrix as that calculated by Klose and Zarembo for the Faddeev-Reshetikhin  model. Furthermore, we found that in spite of the fact that the spectrum of this generalized family of Hamiltonians is real and the S-matrix is unitary, the Hamiltonian is not Hermitian but is PT symmetric. It falls in a class of quantum mechanical systems with a non Hermitian Hamiltonian but unitary S-matrix \cite{Bender:2007nj}.

\section{Strings in $R \times S^3$}

The action for a string in $AdS_5 \times S^5$ is given by 
\begin{equation}
S = \frac{1}{4\pi \alpha^\prime} \int d^2 \sigma (
  G_{\mu\nu}^{(AdS_5)} \partial^\alpha X^\mu \partial_\alpha X^\nu +
  G_{ab}^{(S^5)} \partial^\alpha Y^a \partial_\alpha Y^b ).
\end{equation}
The embedding coordinates in $AdS$ are $X^0, \dots , X^5$ and those for $S^5$ are $Y^1, \dots, Y^6$ and they satisfy 
\begin{equation}
(X^0)^2 + (X^1)^2 + \dots + (X^4)^2 - (X^5)^2 = R^2,  
\end{equation}
\begin{equation}
(Y^1)^2 + \dots + (Y^6)^2 = R^2,
\end{equation}
The truncation to $R \times S^3$ is just given by $X^0 = t$, $ Y^5 = Y^6 = 0$ and vanishing the remaining variables in $AdS$.  The Virasoro constraints then reduce to 
\begin{equation}
\Lambda_{\alpha\beta} + \frac{1}{2} \eta_{\alpha\beta}
\Lambda^\gamma_\gamma = 0,
\end{equation}
where 
\begin{equation}
\Lambda_{\alpha \beta} = \delta_{ab} \partial_\alpha Y^a
\partial_\beta Y^b - \partial_\alpha t \partial_\beta t, \quad a,b
= 1 \dots 4.
\end{equation}
To make connection with the SU(2) principal chiral sigma model we now use the 
standard map of $S^3$ to $SU(2)$
\begin{equation}
g = \left( 
\begin{array}{ccc}
Y^1 + i Y^2 & Y^3 + i Y^4 \\
-Y^3 + i Y^4 & Y^1 - i Y^2
\end{array} 
\right) =  \left( 
\begin{array}{ccc}
z_1 & z_2 \\ - \overline{z}_2 & \overline{z}_1
\end{array}
\right),
\end{equation}
where $\det g = 1$ thanks to the constraint on the embedding coordinates of $S^5$. We then define the $SU(2)$ current in the usual way $J = g^{-1} dg$ so that the string equations of motion are 
\begin{equation}
\partial_\alpha J^\alpha = 0, \qquad \partial^2 t = 0,
\end{equation}
while the Virasoro constraints reduce to 
\begin{equation}
Tr ( J^2_\pm) = -2 (\partial_\pm t)^2.
\end{equation}
We can also choose the gauge $t = \kappa \tau$ so that the Virasoro constraints simplify to $Tr(J^2_\pm) = - 2 \kappa^2$. 

To make contact with the usual form of the Faddeev-Reshetikhin model we recall that $J_\pm$ is a two by two matrix and can be expanded in terms of the Pauli matrices. We do that by introducing the variables $\vec{S}_\pm$ as 
\begin{equation}
J_\pm = i  \kappa \vec{S}_\pm \cdot \vec{\sigma}. 
\end{equation}
Then the Virasoro constraints become $\vec{S}^2_\pm = 1$ and the equations of motion reduce to 
\begin{equation}
\partial_\mp \vec{S}_\pm \mp \kappa  \vec{S}_+ \times \vec{S}_- = 0,
\end{equation}
which is the familiar form of the Faddeev-Reshetikhin equations. As remarked in
the introduction, this model has been solved using the inverse scattering method
\cite{Faddeev:1985qu} and field theoretic techniques \cite{Klose:2006dd}. Our
approach \cite{Das:2007tb} is the diagonalization of the system. 

The Faddeev-Reshetikhin equations can be obtained from the following action
\begin{equation} 
S = \int d^2 x [ C_+(\vec{S}_-) + C_-(\vec{S}_+) + \frac{\kappa}{2}
  \vec{S}_+ \cdot \vec{S}_- ],
\end{equation}
where $C_\pm(\vec{S}_\mp)$ are the usual Wess-Zumino terms
\begin{equation}
C_\pm(\vec{S}_\mp) = -\frac{1}{2} \int_0^1 d\xi \epsilon^{abc} S^a_\mp \partial_\xi S^b_\mp S^c_\mp.
\end{equation}
To get a canonical kinetic term and unconstrained
variables we change variables again to 
\begin{equation}
\phi_\pm = \frac{S^1_\pm + i S^2_\pm }{\sqrt{2}\sqrt{1 + S^3_\pm}},
  \quad S^3_\pm = 1 - 2 | \phi_\pm |^2,
\end{equation}
so that the action takes the form 
\begin{eqnarray}
S &=& \int d^2x \,\, [ \frac{i}{2}(\phi^*_- \partial_+ \phi_- - 
  \phi_- \partial_+ \phi^*_-) + \frac{i}{2}(\phi^*_+ \partial_- \phi_+ -
  \phi_+ \partial_- \phi^*_+) + \kappa ( |\phi_+|^2 + |\phi_-|^2)
\nonumber \\
& - &   \kappa \sqrt{(1 - |\phi_+|^2) (1 - |\phi_-|^2)}(\phi^*_+ \phi_- +
  \phi^*_- \phi_+) - 2\kappa |\phi_+|^2 |\phi_-|^2 ].
\end{eqnarray}
It can be rewritten in a more compact Dirac like form as 
\begin{equation}
S = \int d^2x [ \overline{\phi} i \gamma^\mu D_\mu \phi - m
  \overline{\phi} \phi - g ( \overline \phi \gamma^\mu \phi )^2 +
	   {\cal O}(\phi^6) ],
\end{equation}
where 
\begin{equation} 
\phi = \left(
\begin{array}{c}
\phi_- \\ \phi_+
\end{array}
\right),
\end{equation}
and $D_0 = \partial_0 - im - i g \overline{\phi} \phi$, $D_1 = \partial_1$; $m = \kappa$ and $g =
\kappa/2$. Even though the action looks Lorentz invariant it is not so because of the derivatives. It is just convenient to write the action in this way. Notice also that the square root term was expanded up to fourth order in the fields. 

\section{Diagonalization of the Hamiltonian}

To obtain the two particle S-matrix we need the Hamiltonian up to fourth order in the fields. From the action in the previous section we easily compute the Hamiltonian  
\begin{equation}
\label{17}
 H = \int dx :[ - \overline{\phi} i \gamma^1 \partial_x \phi + m
  \overline{\phi} \phi + g ( \overline{\phi} \gamma^0 \phi \,\, 
  \overline{\phi} \phi) - ( \overline{\phi} \gamma^\mu \phi)^2 ) ]:.
\end{equation}
We assume the standard commutation relations $[\phi_\alpha(x), \phi^\dagger_\beta(y)] = \delta_{\alpha\beta}
  \delta(x-y)$ and choose the false vacuum $\phi(x) |0> = 0$. We can go the true vacuum later on by filling all the negative energy states. 

The two particle states can be written as 
\begin{equation}
|k_1,k_2> = \int dx_1 \, dx_2 \,\, \chi_{\alpha\beta}(x_1,x_2,k_1,k_2) \,\,
\phi^\dagger_\alpha(x_1) \phi^\dagger_\beta(x_2) |0>,
\end{equation}
where $\chi_{\alpha\beta}(x_1,x_2) = \chi_{\beta\alpha}(x_2,x_1)$ is the wave function. For the free theory we can make use of the Dirac technology: define the two component solutions of the momentum space Dirac equation $u(k)$ and $v(k)$ as
\begin{equation}
 ( \slash \!\! k - m) u(k) = 0, \qquad (\slash \!\! k + m) v(k) = 0,
\end{equation}
with 
\begin{equation}
u(k) = \sqrt{m} \left( 
\begin{array}{c}
e^{\beta/2} \\
e^{-\beta/2}
\end{array} \right),
\qquad
v(k) = \sqrt{m} \left( 
\begin{array}{c}
-e^{\beta/2} \\
e^{-\beta/2}
\end{array} \right),
\end{equation}
where $\beta$ is the rapidity defined as usual by $k^0 = m \cosh \beta$ and $k^1 = m \sinh\beta$. We now look for eigenfunctions of the momentum with eigenvalue $k_1 + k_2$ and eigenfunctions of the Hamiltonian with eigenvalue $ E = \sqrt{m^2 + k_1^2} + \sqrt{m^2 + k_2^2}$. They can be easily found to be 
\begin{equation}
\chi^{(\pm)}_{\alpha\beta}(x_1,x_2,k_1,k_2) = e^{i(k_1 x_1 + k_2
  x_2)} u_\alpha(k_1) u_\beta(k_2) \pm  e^{i(k_1 x_2 + k_2
  x_1)} u_\alpha(k_2) u_\beta(k_1).
\end{equation}

We now turn to the interacting case. The ansatz is a superposition of the free solutions with the same eigenvalues for the momentum and energy as in the free case as usual in integrable models. We denote it by 
\begin{equation}
\chi_{\alpha\beta}(x_1,x_2) = \chi_{\alpha\beta}^{(+)}(x_1,x_2) +
\lambda(k_1,k_2) \,\, \epsilon(x_1 - x_2) \,\,
\chi_{\alpha\beta}^{(-)}(x_1,x_2),
\end{equation}
with $\lambda(k_1,k_2)$ to be determined. Written in this form the two particle S-matrix takes the form $S = (1 - \lambda)/(1 + \lambda)$. Remarkably, there is no solution for the eigenvalue problem! The Hamiltonian is not diagonalizable.

Let us take a closer look at the interaction Hamiltonian which can be written as 
\begin{equation}
H_I |k_1,k_2>  = \frac{g}{2} \int dx_1 \,\, dx_2 \,\, \delta(x_1 -
x_2) V_{\alpha\beta\gamma\delta} \,\,  \chi_{\gamma\delta}(x_1,x_2)  \,\, 
\phi^\dagger_\alpha(x_1) \phi^\dagger_\beta(x_2) |0>,
\end{equation}
showing that the interaction is produced by a single Dirac delta interaction. Instead of solving the eigenvalue problem we could try to match the discontinuity produced by the delta function. Again, no solution was found. If the Lorentz violating term involving the $\gamma^0$ contribution in (\ref{17}) is eliminated then a solution can be found. Removing the $\gamma^0$ term in the Hamiltonian reduces it to the known bosonic massive Thirring model and the solution found corresponds indeed to the two particle S-matrix of this relativistic model 
\begin{equation}
S = \frac{1 - i g \coth(\beta_1 - \beta_2)/2}{1 + ig \coth(\beta_1 -
  \beta_2)/2}.
\end{equation}

To get the S-matrix for the Faddeev-Reshetikhin model, which is not
relativistic, we need only matrix elements between say positive-positive energy
states. We do not need the knowledge of the intermediate states. So consider
again the eigenvalue equation and take the inner product with a positive energy
state and solve for $\lambda(k_1,k_2)$. Now a solution exists and we obtain
\begin{equation}
\lambda = i g \frac{ \cosh{\frac{\beta_1 + \beta_2}{2}} - \cosh{\frac{\beta_1
      - \beta_2}{2}} }{\sinh{\frac{\beta_1 - \beta_2}{2}}},
\end{equation}
which is indeed the correct form for $\lambda$ to produce the two particle S-matrix for the Faddeev-Reshetikhin model. So, a diagonalizable Hamiltonian  leads to a S-matrix but having a S-matrix does not automatically imply that the Hamiltonian is diagonalizable!

\section{General Quartic Hamiltonian}

At this point is natural to look for the most general quartic Hamiltonian which can be diagonalized, under the assumption of integrability, and its corresponding two particle S-matrix. We found a one parameter set of models 
\begin{equation}
\label{H}
H_I = \int dx \left[ \alpha ( \overline{\phi} \gamma^\mu \phi )^2 + \beta
  \overline{\phi} \gamma^\mu \gamma^0 \phi \,\, \overline{\phi}
  \gamma_\mu \phi \right],
\end{equation}
with 
\begin{equation}
\lambda = - i \left[ \alpha  \coth \frac{\beta_1 - \beta_2}{2} +
  \beta \frac{\cosh\frac{\beta_1 + \beta_2}{2}}{\sinh\frac{\beta_1 -
      \eta_2}{2}} \right], \qquad S = \frac{1 - \lambda}{1 + \lambda}.
\end{equation}
For $\alpha=g$ and $\beta=0$ we regain the bosonic massive Thirring model
discussed in the previous section. For $\alpha = -\beta = g$ we get a one
parameter family of models with the same two particle S-matrix as the
Faddeev-Reshetikhin model. Due to the Fierz identity
\begin{equation}
 (\overline{\phi} \phi)^2 - (\overline{\phi} \gamma_5 \phi)^2 = (\overline{\phi}
\gamma^\mu \phi)^2,
\end{equation}
the Hamiltonian (\ref{H}) with $\beta=0$ reduces to the bosonic chiral
Gross-Neveu
model and it shares the same S-matrix as the bosonic massive Thirring model. As
far as we know, this has not been noticed before. This was verified explicitely 

Notice that hermiticity is in trouble in the $\beta$ term since 
$(\overline{\phi}\gamma^0 \gamma^1 \phi)^\dagger = - \overline{\phi}\gamma^0
\gamma^1 \phi $ is pure 
  imaginary while $(\overline{\phi} \gamma^1 \phi)^\dagger = \overline{\phi} \gamma^1 \phi$ is real. Then the  Hamiltonian is not hermitian for real $\beta$ but its S-matrix is unitary!
 Quantum mechanical theories with non-Hermitian Hamiltonians which still have
real energy are well known \cite{Bender:2007nj}. For these theories the
requirement of hermiticity is replaced by the weaker requirement of being PT
symmetric. In our case, the PT symmetry is given by 
\begin{eqnarray}
P: &\quad& \phi(x,t) \rightarrow \eta_P \gamma^0 \phi(-x,t), \nonumber \\
T: &\quad& \phi(x,t) \rightarrow \eta_T C \gamma^0 \gamma^1 \phi(x,-t),
\end{eqnarray}
and the Hamiltonian is PT symmetric if $\alpha$ and $\beta$ are real. This is possibly the first example of an integrable field theory which has a non-Hermitian Hamiltonian, is PT symmetric, and has a unitary S-matrix. 

\section{Conclusion}

We have studied in detail the Faddeev-Reshetikhin model which appears when string theory in $AdS_5 \times S^5$ is truncated to $R \times S^3$. Even though the S-matrix
of the theory has been calculated using field theoretic methods, diagonalization of the
Hamiltonian is essential to carry out the Bethe ansatz analysis. We found that the quartic
Hamiltonian for this model is not diagonalizable in the two particle sector. 
We found that the difficulty is in the term in the interaction Hamiltonian 
violating Lorentz invariance. If we take the inner product of the eigenvalue equation 
with positive energy states the problematic term disappears leading to the correct S-matrix
element calculated earlier. Further investigation shows that the interaction 
generates intermediate states that are orthogonal to the positive energy out states and,
therefore, cannot be observed in the S-matrix calculation but are relevant in the diagonalization of the system. To the best of our
knowledge, this is a new feature that has not been observed earlier in the study of integrable
systems. It follows, therefore, that while the diagonalization of a Hamiltonian leads to the
S-matrix of the theory, the knowledge of the S-matrix element by other means does not automatically guarantee the diagonalizability of the Hamiltonian of
the system. We have also determined the most general Hamiltonian with quartic interactions that
can be diagonalized as well as its S-matrix. Among various special cases, it
also includes a generalized Hamiltonian that can be diagonalized with the same
S-matrix as that 
calculated by Klose and Zarembo. We showed that although this general Hamiltonian leads
to a real spectrum and a unitary S-matrix, it is not Hermitian. However, it 
is PT symmetric and the wave functions are also invariant under PT transformations. As
a result, the theory is in an unbroken phase of PT symmetry which is the reason for the
reality of the spectrum as well as the unitarity of the S-matrix.

\begin{theacknowledgments}
  I would like to thank the organizers of the conference Ten Years of AdS/CFT for the kind invitation to deliver  this talk and to A. Das and A. Melikyan for collaboration in this work. This work was supported by CNPq, FAPESP and PROSUL grant No. 490134/2006-8.
\end{theacknowledgments}





 


\end{document}